\documentclass{elsart}
\usepackage{graphicx,psfrag,amsmath, epsfig}

\renewcommand{\d}{\partial}

\newcommand{\bold}[1]{{\bf{#1}}}
\newcommand{\mathbold}[1]{\mbox{\boldmath $\bf#1$}}

\newcommand{\pim}{\pi_{\scriptscriptstyle{-}}}
\newcommand{\pin}{\pi_{\scriptscriptstyle{0}}}
\newcommand{\Kp}{K_{\scriptscriptstyle{+}}}

\newcommand{\Kn}{K_{\scriptscriptstyle{0}}}
\newcommand{\Mpic}{M_{\pi^{\scriptscriptstyle{+}}}}
\newcommand{\Mpin}{M_{\pi^{\scriptscriptstyle{0}}}}
\newcommand{\MKc}{M_{\scriptscriptstyle{K^+}}}
\newcommand{\MKn}{M_{\scriptscriptstyle{K^0}}}

\newcommand{\w}[2]{\omega_{#1}(#2)}
\newcommand{\Lagr}{\mathcal{L}}
\newcommand{\Order}[1]{\mathcal{O}#1}
\newcommand{\order}[1]{{\it o}#1}
\newcommand{\nn}{\nonumber}
\newcommand{\sss}{\scriptscriptstyle}

\begin{document}
\begin{frontmatter}
\title{Decay widths and energy shifts of $\mathbold{\pi\pi}$ and $\mathbold{\pi K}$ atoms}
  \author{J. Schweizer}
  \address{Institute for Theoretical Physics, University of Bern,\\Sidlerstrasse
    5, CH-3012 Bern, Switzerland\\E-mail: schweizer@itp.unibe.ch}
\begin{abstract}
We calculate the S-wave decay widths and energy shifts for $\pi^+\pi^-$ and
$\pi^\pm K^\mp$ atoms in the framework of QCD $+$ QED. The evaluation - valid at next-to-leading order in isospin symmetry
breaking - is performed within a non-relativistic effective
field theory. The results are of interest for future hadronic atom experiments.
\end{abstract}
\end{frontmatter}
\section{Introduction}
\label{section: Introduction}
Nearly fifty years ago, Deser {\it et al.} \cite{Deser:1954vq} derived the
formulae for the decay width and strong energy shift of pionic hydrogen at
leading order in isospin symmetry breaking. Similar relations also hold for
$\pi^+\pi^-$ \cite{Palfrey:kt} and $\pi^- K^+$
atoms, which decay predominantly into $2\pi^0$ and $\pi^0 K^0$, respectively. These
Deser-type relations allow to extract the scattering lengths from 
measurements of the decay width and the strong energy shift. The DIRAC
collaboration \cite{Adeva:1994xz} at CERN intends to measure the lifetime of
pionium in its ground state at the $10\%$ level, which will allow to extract the scattering length difference $|a_0^0-a_0^2|$ at $5\%$ accuracy. The
experimental result can then be compared with theoretical predictions for
the S-wave scattering lengths
\cite{Weinberg:1966kf,Colangelo:2000jc,Colangelo:2001df}
and with the results from other experiments \cite{Rosselet:1976pu}. Particularly
interesting is the fact that one may determine in this manner the nature of
the SU(2)$\times$SU(2) spontaneous chiral symmetry breaking experimentally
\cite{Knecht:1995tr}. New experiments are proposed for CERN PS and J-PARC in Japan \cite{Adeva:2000vb}. In order to determine the scattering lengths from such experiments, the
theoretical expressions for the decay width and the strong energy shift must
be known to an accuracy that matches the experimental precision. For this
reason, the ground state decay width of pionium has been evaluated at
next-to-leading order \cite{pipi,Jallouli:1997ux,Labelle:1998gh,Gasser:1999vf,Eiras:2000rh,Gasser:2001un} in the isospin symmetry breaking parameter $\delta$, where both the
  fine-structure constant $\alpha$ and $(m_u-m_d)^2$ count as
  $\Order{(\delta)}$. The aim of the present article is to provide the
  corresponding formulae for the S-wave decay widths and strong energy shifts
  of pionium and the $\pi^\pm K^\mp$ atom at next-to-leading order in isospin
  symmetry breaking. A detailed derivation of the results will be provided
  elsewhere \cite{Schweizer}. The strong energy shift of the
  $\pi^\pm K^\mp$ atom is proportional to the sum of the isospin even and odd S-wave $\pi K$ scattering lengths $a_0^+ + a_0^-$. This sum \cite{Bernard:1990kx,Roessl:1999iu,Nehme:2001wa,Kubis,Buettiker:2003pp} is sensitive to the combination of low-energy constants
  $2L_6^r+L_8^r$ \cite{Gasser:1984gg}. The consequences of this observation for the SU(3)$\times$SU(3) quark condensate \cite{Descotes-Genon:1999uh} remain to be worked out.

\section{Non-relativistic framework}
The non-relativistic effective Lagrangian framework has proven to be
a very efficient method to investigate bound state characteristics
\cite{Labelle:1998gh,Gasser:2001un,Caswell:1985ui,Lyubovitskij:2000kk}. The non-relativistic Lagrangian is exclusively determined by symmetries, which are
rotational invariance, parity and time reversal. It provides a systematic
expansion in powers of the isospin breaking parameter $\delta$. What concerns
the $\pi^- K^+$ atom, we count both
$\alpha$ and $m_u-m_d$ as order $\delta$. The different power counting for the
$\pi^+\pi^-$ and $\pi^- K^+$ atoms are due to the fact that in QCD,
the chiral expansion of the pion mass difference $\Delta_\pi=\Mpic^2-\Mpin^2$
is of second order in $m_u-m_d$, while the kaon mass difference
$\Delta_K=\MKc^2-\MKn^2$ starts at first order in $m_u-m_d$. In the sector
with one or two mesons, the non-relativistic $\pi K$ Lagrangian is $\Lagr_{\sss \rm NR} = \Lagr_1+\Lagr_2$. The first term contains the one-pion and one-kaon sectors,
\begin{eqnarray}
  \Lagr_1 &=& \frac{1}{2}(\bold{E}^2-\bold{B}^2)+h_0^\dagger \Big( i \d_t
  -M_{h^0}+\frac{\Delta}{2M_{h^0}}+\frac{\Delta^2}{8M_{h^0}^3}+\cdots\Big)h_0\nn\\
&&+ \sum_{\pm}h_\pm^\dagger \Big( i D_t
  -M_{h^+}+\frac{\bold{D}^2}{2M_{h^+}}+\frac{\bold{D}^4}{8M_{h^+}^3}+\cdots\Big)h_\pm,
\label{freeLagr}
\end{eqnarray}
where $\bold{E} = -\nabla A_0-\dot{\bold{A}}$, $\bold{B}= \nabla\times
\bold{A}$ and the quantity $h=\pi,K$ stands for the
non-relativistic pion and kaon fields. We work in the Coulomb gauge and eliminate the $A^0$ component of the photon
field by the use of the equations of motion. The covariant derivatives are
given by $D_t h_\pm = \d_t h_\pm \mp i e A_0 h_\pm$ and $\bold{D}h_\pm = \nabla
  h_\pm \pm i e \bold{A}h_\pm$, where $e$ denotes the electromagnetic coupling. What concerns the one-pion-one-kaon sector, we only list
  the terms needed to evaluate the decay width and the energy shift of the
  $\pi^-K^+$ atom at order $\delta^{9/2}$ and $\delta^4$, respectively,
\begin{equation}
  \Lagr_2= C_1'\pim^\dagger\Kp^\dagger\pim\Kp +C_2\left(
    \pim^\dagger\Kp^\dagger\pin\Kn+\textrm{h.c}\right)+
    C_3\pin^\dagger\Kn^\dagger\pin\Kn+\cdots
\label{Lagr}
\end{equation}
 The ellipsis stands for
higher order terms\footnote{The basis of operators containing two space derivatives can
  be chosen such that none of them contributes to the energy
  shift and decay width at next-to-leading order in isospin symmetry breaking
  \cite{Schweizer}.}. We work in the center of mass system and thus omit terms
proportional to the total 3-momentum. The total and reduced masses read  
\begin{equation}
  \Sigma_i = M_{\pi^i}+M_{\sss K^i}, \quad
   \mu_i = \frac{M_{\pi^i}M_{\sss K^i}}{M_{\pi^i}+M_{\sss K^i}}, \quad i=+,0. 
\end{equation}
The coupling constant $C_1'$ contains contributions coming from the
electromagnetic form factors of the pion and kaon,
\begin{equation}
  C_1' = C_1- e^2\lambda,\quad \lambda =\frac{1}{6}\left( \langle
  r^2_{\pi^+} \rangle+\langle r^2_{\sss K^+} \rangle\right),
\end{equation}
where $\langle
  r^2_{\pi^+} \rangle$ and $\langle r^2_{\sss K^+} \rangle$ denote the charge radii of the charged pion
and kaon, respectively. The low energy constants $C_1,\dots,C_3$ may be determined through matching the $\pi K$ amplitude at threshold for
various channels, see section \ref{sec: matching}.

To evaluate the energy shift and decay width of the $\pi^- K^+$ atom at
next-to-leading order in isospin symmetry breaking, we make use of
resolvents. For a detailed discussion of the technique, we refer to
Ref. \cite{Gasser:2001un}. Here, we simply list the results. We use
dimensional regularization, to treat both ultraviolet and infrared singularities. Up to and including order $\delta^{9/2}$, the decay into $\pi^0 K^0$ is the
only decay channel contributing, and we get for the total S-wave decay width
\begin{eqnarray}
 \Gamma_{n} &=& \frac{\alpha^3\mu_+^3}{n^3\pi^2}\mu_0 k_0
 C_2^2\left[1-\frac{\mu_0^2k_0^2C_3^2}{4\pi^2}-\frac{\alpha\mu_+^2
 C_1\xi_n}{\pi}\right.\nn\\
&&\left.+\frac{5\mu_0 k^2_0}{8}\frac{\Mpin^3+\MKn^3}{\Mpin^3
 \MKn^3}\right]+\Order{(\delta^{5})},
\label{decaywidth}
\end{eqnarray}
where $k_0 = [2\mu_0(\Sigma_+-\Sigma_0-\alpha^2\mu_+/(2n^2))]^{1/2}$ is of
order $\delta^{1/2}$. The function $\xi_n$ develops an
ultraviolet singularity as $d\rightarrow3$,
\begin{eqnarray}
  \xi_n &=& \Lambda(\mu)-1+2\left[{\rm ln}\frac{\alpha}{n}+{\rm
  ln}\frac{2\mu_+}{\mu}+\psi(n)-\psi(1)-\frac{1}{n}\right],\nn\\
  \Lambda(\mu) &=&\mu^{2(d-3)}\left[\frac{1}{d-3}-{\rm
    ln}4\pi-\Gamma'(1)\right],
\end{eqnarray}
with $\psi(n)=\Gamma'(n)/\Gamma(n)$ and the running scale $\mu$. At order $\delta^4$, the total energy shift may be split into a strong part and an electromagnetic part, according to
\begin{equation}
  \Delta E_{n} = \Delta E^{\rm h}_{n}+\Delta E^{\rm em}_{n}.
\label{DeltaEtot}
\end{equation}
For the discussion of the electromagnetic energy shift, we refer to section
\ref{sec: results}. The strong S-wave energy shift reads
\begin{equation}
   \Delta E^{\rm h}_{n} = -\frac{\alpha^3\mu_+^3}{\pi n^3}\left[ C_1-\frac{\alpha\mu_+^2}{2\pi}C_1^2\xi_n-\frac{\mu_0^2k_0^2}{4\pi^2}C_2^2C_3\right]+\Order{(\delta^5)}.
\label{energyshift}
\end{equation}
The results for the decay width (\ref{decaywidth}) and energy shift (\ref{energyshift}) are valid at next-to-leading
order in isospin symmetry breaking. 
\section{Matching the low-energy constants}
\label{sec: matching}
The coupling constants $C_i$ can be determined through matching the non-relativistic
  and the relativistic amplitudes at threshold. The coupling $C_3$ is needed
  at order $\delta^0$ only. However, we have to determine both
  $C_1$ and $C_2$ at next-to-leading order in isospin symmetry breaking. The relativistic amplitudes are related to the non-relativistic ones through
\begin{equation}
  T^{lm;ik}_{\rm R}(\bold{q};\bold{p})=
  4\left[\w{i}{\bold{p}}\w{k}{\bold{p}}\w{l}{\bold{q}}\w{m}{\bold{q}}\right]^{\frac{1}{2}}
  T^{lm;ik}_{\sss \rm NR}(\bold{q};\bold{p}),
\label{matching}
\end{equation}
with $\w{i}{\bold{p}}=(M_i^2+\bold{p}^2)^{1/2}$. The 3-momentum $\bold{p}$
denotes the center of mass momentum of the incoming particles, $\bold{q}$ the
one of the outgoing particles. The effective Lagrangian
in Eqs. (\ref{freeLagr}) and (\ref{Lagr}), allows us to evaluate the
non-relativistic $\pi^- K^+\rightarrow\pi^0 K^0$ and $\pi^- K^+\rightarrow
\pi^- K^+$ scattering amplitudes at threshold at order $\delta$. In the isospin symmetry limit, the effective couplings $C_1$, $C_2$
and $C_3$ are
\begin{equation}
  C_1 = \frac{2\pi}{\mu_+}\left(a_0^++a_0^-\right),\quad
  C_2 = -\frac{2\sqrt{2}\pi}{\mu_+}a_0^-,\quad
  C_3 = \frac{2\pi}{\mu_+}a_0^+,
\label{CI}
\end{equation}
where the S-wave scattering lengths\footnote{$a_0^+$ and $a_0^-$ are normalized as in
  Ref. \cite{Bernard:1990kx}.} $a_0^+=1/3(a_0^{1/2}+2a_0^{3/2})$ and $a_0^-=1/3(a^{1/2}_0-a^{3/2}_0)$ are defined in QCD, at $m_u
  = m_d$ and $M_\pi \doteq\Mpic$, $M_{\sss K}\doteq\MKc$. By substituting
  these relations into the expression for the decay width
(\ref{decaywidth}) and the strong energy shift (\ref{energyshift}), one obtains
the Deser-type formulae \cite{Deser:1954vq,Palfrey:kt}.
\begin{figure}[ttbp]
\begin{center}
\leavevmode
\makebox{\includegraphics[height=2.6cm]{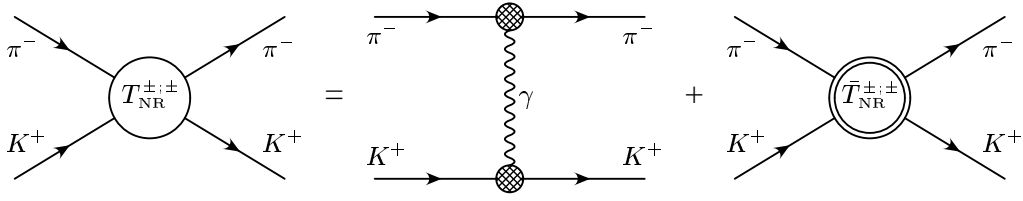}}
\caption{Non-relativistic $\pi^- K^+ \rightarrow \pi^- K^+$ scattering amplitude. The blob
  describes the vector form factor of the pion and
  kaon. $\bar{T}^{\pm;\pm}_{\sss \rm NR}$ denotes the truncated amplitude.}
\label{fig: TpiKel}
\end{center}
\end{figure}
We demonstrate the matching at next-to-leading order in $\delta$ by means of
the $\pi^- K^+\rightarrow \pi^- K^+$ amplitude. In the presence of virtual
photons, we first have to subtract the one-photon exchange diagram from the
full amplitude, as displayed in Fig. \ref{fig: TpiKel}. The coupling constant
$C_1$ is determined by the truncated part $\bar{T}^{\pm;\pm}_{\sss \rm
  NR}$, which contains an infrared singular Coulomb phase $\theta_c$ as $d\rightarrow3$,
\begin{eqnarray}
  \bar{T}^{\pm;\pm}_{\sss \rm NR}(\bold{p}; \bold{p}) &=&
  e^{2i\alpha\theta_c}\hat{T}^{\pm;\pm}_{\sss \rm NR}(\bold{p}; \bold{p}),\nn\\
\theta_c
  &=&\frac{\mu_+}{|\bold{p}|}\mu^{d-3}\left\{\frac{1}{d-3}-\frac{1}{2}\left[{\rm
  ln}4\pi+\Gamma'(1)\right]+{\rm ln}\frac{2|\bold{p}|}{\mu}\right\}.
\end{eqnarray}
At order $\delta$, the remainder $\hat{T}^{\pm;\pm}_{\sss \rm NR}$ is free of
infrared singularities at threshold. The real part of $\hat{T}^{\pm;\pm}_{\sss \rm NR}$ is given by
\begin{equation}
{\rm Re}\,\hat{T}^{\pm;\pm}_{\sss \rm NR}(\bold{p}; \bold{p}) = \frac{B_1'}{|\bold{p}|}+B_2'
  {\rm ln}\frac{|\bold{p}|}{\mu_+}+\frac{1}{4\Mpic\MKc}{\rm Re}\,A^{\pm;\pm}_{{\rm
  thr}}+\Order{(\bold{p})},
\label{ReTthrc}
\end{equation}
with $B_1' = C_1 \alpha \pi\mu_++\order{(\delta)}$, $B_2' = -C_1^2
  \alpha \mu_+^2/\pi+\order{(\delta)}$ and
\begin{eqnarray}
  \frac{1}{4\Mpic\MKc}{\rm Re}\,A^{\pm;\pm}_{{\rm
  thr}} &=& C_1\left\{1+\frac{C_1\alpha\mu_+^2}{2\pi}\left[1-\Lambda(\mu)-{\rm
  ln}\frac{4\mu_+^2}{\mu^2}\right]\right\}\nn\\
&&-\frac{C_2^{2}C_3\mu_0^3}{2\pi^2}(\Sigma_+-\Sigma_0)+\order{(\delta)}.
\label{ReAthrc}
\end{eqnarray}
Here, the ultraviolet pole term $\Lambda(\mu)$ is removed by renormalizing the
coupling $C_1$. The renormalization of $C_1$
eliminates at the same time the ultraviolet divergence contained in the
expression for the energy shift (\ref{energyshift}).
The calculation of the relativistic $\pi^- K^+\rightarrow\pi^-K^+$ scattering amplitude was performed at
$\Order{(p^4,e^2p^2)}$ in
Refs. \cite{Nehme:2001wa,Kubis}. Both the Coulomb phase
 and the logarithmic singularity in Eq. (\ref{ReTthrc}) are absent in the
 real part of the relativistic amplitude at this order of accuracy, they first
 occur at order $e^2p^4$. The quantity ${\rm Re}\,A^{\pm;\pm}_{{\rm thr}}$
 denotes the constant term occurring in the real part of the truncated
 relativistic threshold amplitude. The coupling constant $C_2$ may be
 determined analogously by matching the non-relativistic $\pi^-K^+\rightarrow\pi^0 K^0$ amplitude to the relativistic one at order $\delta$.

\section{Results for the $\mathbold{\pi^- K^+}$ atom}
\label{sec: results}
 The result for the decay width and strong energy shift are valid at
next-to-leading order in isospin symmetry breaking, and to all orders in the
chiral expansion. We get for the decay width at order $\delta^{9/2}$, in terms
of the relativistic $\pi^-K^+\rightarrow\pi^0K^0$ threshold amplitude,
\begin{equation}
  \Gamma_{n} =
  \frac{8}{n^3}\alpha^3\mu_+^2p^*_n\mathcal{A}^2\left(1+K_n\right),\quad
  \mathcal{A} = -\frac{1}{8\sqrt{2}\pi}\frac{1}{\Sigma_+}{\rm
  Re}\,A^{00;\pm}_{\rm thr}+\order{(\delta)},
\end{equation}
where
\begin{eqnarray}
  K_n &=& \frac{\Mpic
  \Delta_K+\MKc\Delta_\pi}{\Mpic+\MKc}(a^+_0)^2\nn\\
&&-4\alpha\mu_+(a^+_0+a^-_0)\left[\psi(n)-\psi(1)-\frac{1}{n}+{\rm
  ln}\frac{\alpha}{n}\right]+\order{(\delta)}.
\end{eqnarray}
The outgoing relative 3-momentum
\begin{equation}
  p^*_n = \frac{1}{2 E_n}\lambda\left(E_n^2,\Mpin^2,\MKn^2\right)^{1/2},
\end{equation}
with $\lambda(x,y,z)=x^2+y^2+z^2-2x y-2x z-2y z$, is chosen such that the total
final state energy corresponds to $E_n=\Sigma_+-\alpha^2\mu_+/(2n^2)$. 
The quantity ${\rm Re}\,A^{00;\pm}_{\rm thr}$ is calculated as follows. One
evaluates the relativistic $\pi^-K^+\rightarrow\pi^0K^0$ amplitude near
threshold and removes the divergent Coulomb phase. The real part contains
singularities $\sim 1/|\bold{p}|$ and $\sim{\rm ln}|\bold{p}|/\mu_+$. The
constant term in this expansion corresponds to ${\rm Re}\,A^{00;\pm}_{\rm
  thr}$. The normalization is chosen such that 
\begin{equation}
  \mathcal{A}=a^-_0+\epsilon.
\label{eq: A}
\end{equation}
The isospin breaking corrections $\epsilon$ have been
evaluated at $\Order{(p^4,e^2p^2)}$ in Refs. \cite{Kubis,Kubis:2001ij}. See
also the comments in section \ref{sec: numerics}.

We now discuss the various energy shift contributions. According to Eq. (\ref{DeltaEtot}),
the energy shift at order $\delta^4$ is split into an electromagnetic part
$\Delta E^{\rm em}_{n}$ and the strong part $\Delta E^{\rm h}_{n}$ in Eq. (\ref{energyshift}). The electromagnetic energy shift
contains both pure QED corrections as well as finite size effects due to the
charge radii of the pion and kaon, contained in $\lambda$. The pure
electromagnetic corrections have been evaluated in Ref. \cite{Nandy:rj} for arbitrary angular
momentum $l$. We checked\footnote{We thank A. Rusetsky for a very useful
  communication concerning technical aspects of the calculation.} that the
electromagnetic energy shift at order $\alpha^4$ indeed amounts to
\begin{eqnarray}
   \Delta E^{\rm em}_{nl}
   &=&\frac{\alpha^4\mu_+}{n^3}\left(1-\frac{3\mu_+}{\Sigma_+}\right)\left[\frac{3}{8n}-\frac{1}{2l+1}\right]+\frac{4\alpha^4\mu_+^3\lambda}{n^3}\delta_{l0}\nn\\
&& +\frac{\alpha^4\mu_+^2}{\Sigma_+}\left[\frac{1}{n^3}\delta_{l0}+\frac{1}{n^4}-\frac{3}{n^3(2l+1)}\right]+\Order{(\alpha^5{\rm
   ln}\alpha)}.
\label{eq: DeltaEem}
\end{eqnarray}
Here, the first term is generated by the mass insertions, the second contains
the finite size effects and the last stems from the one-photon exchange
contribution. The strong S-wave energy shift reads at order $\delta^4$,
\begin{equation}
  \Delta E^{\rm h}_{n} =
  -\frac{2\alpha^3\mu_+^2}{n^3}\mathcal{A}'\left(1+K_n'\right),\quad
\mathcal{A}'=\frac{1}{8\pi\Sigma_+}{\rm Re}\,A^{\pm;\pm}_{{\rm
  thr}}+\order{(\delta)},
\label{DeltaE}
\end{equation}
with
\begin{equation}
K_n' = -2\alpha\mu_+(a_0^++a_0^-)\left[\psi(n)-\psi(1)-\frac{1}{n}+{\rm
  ln}\frac{\alpha}{n}\right]+\order{(\delta)}.
\end{equation}
In the isospin limit, the
normalized relativistic amplitude
\begin{equation}
  \mathcal{A}'=a^+_0+a^-_0+\epsilon',
\label{eq: Ap}
\end{equation}
 reduces to the sum of the isospin even and odd scattering lengths. The corrections
$\epsilon'$ have been obtained at $\Order{(p^4,e^2p^2)}$ in Refs. \cite{Nehme:2001wa,Kubis}. See
also the comments in section \ref{sec: numerics}. The result for $\Delta E^{\rm h}_{1}$ in Eq. (\ref{DeltaE}) agrees with the one obtained for the strong energy shift of the ground state in pionic hydrogen
\cite{Lyubovitskij:2000kk}, if we replace $\mu_+$ with the reduced
mass of the $\pi^- p$ atom and ${\rm Re}\,A^{\pm;\pm}_{{\rm
  thr}}$ with the constant term in the threshold expansion for the real
part of the truncated $\pi^- p\rightarrow \pi^- p$ amplitude. 

What remains to be added are the vacuum polarization contributions \cite{Eiras:2000rh,vacpol}, which are
formally of higher order in $\alpha$, however numerically not negligible. The
vacuum polarization leads to an energy level shift $\Delta E^{\rm vac}_{nl}$ as
well as to a change in the Coulomb wave function of the $\pi^- K^+$ atom at the
origin $\delta\psi_{{\sss K}, n}(0)$. For the first two energy levels, $\Delta
E^{\rm vac}_{nl}$ \cite{Eiras:2000rh,vacpol} is given numerically in table \ref{table: numericsK},
 section \ref{sec: numerics}. Formally of
order $\alpha^{2l+5}$, this contribution is enhanced due to its large coefficient
containing $(\mu_+/m_e)^{2l+2}$. The modified Coulomb wave function affects both,
the decay width and the strong energy shift, see section \ref{sec: numerics}. 

As discussed in section \ref{sec: numerics}, the electromagnetic contributions (\ref{eq: DeltaEem}) are known to a high
precision. Further, the strong shift in the
$n$P state is very much suppressed (order $\alpha^5$). A future
measurement of the energy splitting between the $n$S and $n$P states will
therefore allow to extract the strong S-wave energy shift in Eq. (\ref{DeltaE}), and to determine
the combination $a_0^+ + a_0^-$ of the $\pi K$ scattering lengths. The energy
splitting between the 2S and 2P states is given by
\begin{eqnarray}
  \Delta E_{2s-2p} &=& \Delta E_2^{\rm h}+\Delta E_{20}^{\rm em}-\Delta
  E_{21}^{\rm em}+\Delta E_{20}^{\rm vac}-\Delta E_{21}^{\rm vac}\nn\\
&=&-1.4 \pm 0.1\,{\rm eV}.
\end{eqnarray}
The uncertainty displayed is the one in $\Delta E_{2}^{\rm h}$ only. For the numerical values of the various energy shift contributions, see table
  \ref{table: numericsK} in section \ref{sec: numerics}.

\section{Results for pionium}
The decay rate and strong energy shift of pionium can be obtained from the
formulae in Eqs. (\ref{decaywidth}) and (\ref{energyshift}) through the following
substitutions of the masses $\MKc\rightarrow \Mpic$, $\MKn\rightarrow \Mpin$
and the coupling constants $C_1\rightarrow c_1$, $C_2\rightarrow \sqrt{2}(c_2-2c_4
\Delta_\pi)$ and $C_3\rightarrow 2c_3$ \cite{Schweizer}. The $c_i$ are the
low-energy constants defined in Ref.~\cite{Gasser:2001un}.
The S-wave decay width of the $\pi^+\pi^-$ atom reads at order $\delta^{9/2}$,
in terms of the relativistic $\pi^+\pi^- \rightarrow \pi^0\pi^0$ threshold amplitude,
\begin{eqnarray}
  \Gamma_{\pi, n} &=&
  \frac{2}{9n^3}\alpha^3p^*_{\pi, n}\mathcal{A}_\pi^2\left(1+K_{\pi,
  n}\right),\nn\\
\mathcal{A}_\pi &=& a_0^0-a_0^2+\epsilon_\pi\nn,\\
 K_{\pi, n} &=&
  \frac{\kappa}{9}\left(a_0^0+2a_0^2\right)^2-\frac{2\alpha}{3}\left(2a_0^0+a_0^2\right)\left[\psi(n)-\psi(1)-\frac{1}{n}+{\rm
  ln}\frac{\alpha}{n}\right]+\order{(\delta)},\nn\\
p^*_{\pi, n}&=&\left(\Delta_\pi-\frac{\alpha^2}{4n^2}\Mpic^2\right)^{1/2},
\label{Gammapi0pi0}
\end{eqnarray}
where $\kappa =\Mpic^2/\Mpin^2-1$. The quantity $\mathcal{A}_\pi$ is defined
  as in Refs. \cite{Gasser:1999vf,Gasser:2001un}. The isospin
  symmetry breaking corrections $\epsilon_\pi$ have been evaluated at $\Order{(p^4, p^2e^2)}$ in
  Refs. \cite{Gasser:1999vf,Gasser:2001un,Knecht:1997jw}. For the decay width of the ground
  state at order $\delta^{9/2}$, we reproduce the result obtained in
  Refs. \cite{Jallouli:1997ux,Gasser:1999vf,Gasser:2001un}. The
  electromagnetic energy shift $\Delta E^{\rm em}_{\pi, nl}$
  is obtained from Eq. (\ref{eq:
  DeltaEem}) through the above mass substitutions and $\lambda\rightarrow
  1/3\langle r_{\pi^+}^2 \rangle$. Finally, the S-wave energy shift of the
  $\pi^+\pi^-$ atom reads at order $\delta^4$, in terms of the
  relativistic one-particle irreducible $\pi^+\pi^-\rightarrow \pi^+\pi^-$ amplitude at threshold,
\begin{eqnarray}
  \Delta E_{\pi, n}^{\rm h} &=&
  -\frac{\alpha^3\Mpic}{n^3}\mathcal{A}_\pi'\left(1+K_{\pi, n}'\right),\nn\\
\mathcal{A}_\pi'&=&\frac{1}{6}\left(2a_0^0+a_0^2\right)+\epsilon_\pi',\nn\\
K_{\pi, n}' &=& -\frac{\alpha}{3}\left(2a_0^0+a_0^2\right)\left[\psi(n)-\psi(1)-\frac{1}{n}+{\rm
  ln}\frac{\alpha}{n}\right]+\order{(\delta)},
\label{DeltaEpi}
\end{eqnarray}
where $\mathcal{A}_\pi'$ is defined analogously to the quantity $\mathcal{A}'$ discussed in section \ref{sec: results}. 
 The isospin symmetry breaking contributions $\epsilon_\pi'$ have been
 calculated at $\Order{(e^2p^2)}$ in
 Refs. \cite{Meissner:1997fa,Knecht:2002gz}. For pionium the energy splitting between the 2S and 2P states reads
\begin{eqnarray}
  \Delta E_{\pi, 2s-2p} &=& \Delta E_{\pi, 2}^{\rm h}+\Delta E_{\pi, 20}^{\rm
  em}-\Delta E_{\pi, 21}^{\rm em}+\Delta E_{\pi, 20}^{\rm vac}-\Delta E_{\pi,
  21}^{\rm vac}\nn\\
&=&-0.59\pm 0.01\,{\rm eV}.
\end{eqnarray}
Again the uncertainty displayed is the one in $\Delta E_{\pi, 2}^{\rm h}$ only. The numerical values for the various energy shifts are listed in table
  \ref{table: numericspi}, section \ref{sec: numerics}.

\section{Numerical analysis}
\label{sec: numerics}
\begin{table}[t]
\begin{center}
\begin{tabular}{|l|r|r|r|}\hline
& $\delta_{h, 1}$ & $\delta_{h, 1}'$ & $\delta_{h, 2}'$\\\hline
$\pi^+\pi^-$ atom &$(5.8\pm 1.2)\cdot 10^{-2}$ &$(6.2\pm 1.2)\cdot 10^{-2}$ &$(6.1\pm 1.2)\cdot 10^{-2}$ \\
$\pi^\pm K^\mp$ atom &$(4.0\pm2.2)\cdot 10^{-2}$ &$(1.7\pm2.2)\cdot 10^{-2}$ &$(1.5\pm2.2)\cdot 10^{-2}$ \\\hline
\end{tabular}
\end{center}
 \medskip
 \caption{Next-to-leading order corrections to the Deser-type formulae. \label{table: delta}}
\end{table}

For the S-wave $\pi\pi$ scattering lengths, we use the chiral predictions $a_0^0=
0.220\pm0.005$ and $a_0^2 = -0.0444\pm0.0010$ \cite{Colangelo:2000jc,Colangelo:2001df}. The
correlation matrix for $a_0^0$ and $a_0^2$ is given in Ref. \cite{Colangelo:2001df}. For the isospin symmetry breaking corrections to the $\pi\pi$ threshold
amplitudes (\ref{Gammapi0pi0}) and (\ref{DeltaEpi}), we use $\epsilon_\pi=(0.61\pm0.16)\cdot 10^{-2}$ and $\epsilon_\pi'=
(0.37\pm0.08)\cdot 10^{-2}$ as given in Ref. \cite{Gasser:2001un} and
\cite{Knecht:2002gz}, respectively. For the $\pi K$ scattering lengths, we use the values
from the recent analysis of data and Roy-Steiner equations \cite{Buettiker:2003pp}, $a_0^+=(0.045\pm
0.012)\Mpic^{-1}$ and $a_0^-=(0.090\pm 0.005)\Mpic^{-1}$. The correlation
parameter for $a_0^+$ and $a_0^-$ is given in Ref. \cite{Buettiker:2003pp}.
The isospin breaking corrections to the $\pi K$ threshold amplitudes (\ref{eq:
  A}) and (\ref{eq: Ap}) have been
worked out in \cite{Nehme:2001wa,Kubis,Kubis:2001ij}. Whereas the analytic expressions for
$\epsilon$ and $\epsilon'$ obtained in \cite{Nehme:2001wa,Kubis,Kubis:2001ij} are not identical,
the numerical values agree within the uncertainties quoted in \cite{Kubis}. In the following, we use \cite{Kubis} $\epsilon=(0.1\pm0.1)\cdot
10^{-2}M_{\pi^+}^{-1}$ and
  $\epsilon'=(0.1\pm0.3)\cdot10^{-2}M_{\pi^+}^{-1}$. For the charge radii of
  the pion and kaon, we take $\langle r^2_{\pi^+}\rangle =
(0.452\pm0.013)\,{\rm fm^2}$ and $\langle r^2_{\sss K^+}\rangle
=(0.363\pm0.072)\,{\rm fm^2}$ \cite{Bijnens:2002hp}. 

We obtain for the decay width of the ground state,
\begin{equation}
  \Gamma_{1} =
  8\alpha^3\mu_+^2p_1^*(a_0^-)^2\left(1+\delta_{{\sss K}, 1}\right),\quad
  \Gamma_{\pi, 1} = \frac{2\alpha^3}{9}p_{\pi, 1}^*(a_0^0-a_0^2)^2\left(1+\delta_{\pi, 1}\right),
\label{Gammanum}
\end{equation}
where the corrections $\delta_{h, 1}$, $h=\pi, K$ are given in table \ref{table: delta}. The strong energy shift reads
\begin{eqnarray}
  \Delta E_{n}^{\rm h} &=&
  -\frac{2\alpha^3\mu_+^2}{n^3}(a_0^++a_0^-)\left(1+\delta_{{\sss K}, n}'\right),\nn\\
\Delta E_{\pi, n}^{\rm h}
  &=&-\frac{\alpha^3\Mpic}{6n^3}(2a_0^0+a_0^2)\left(1+\delta_{\pi, n}'\right).
\label{DeltaEnum}
\end{eqnarray}
For the first two energy levels, the corrections $\delta_{h, n}'$ are specified in table
\ref{table: delta}. As mentioned in section \ref{sec: results}, these
corrections to the Deser-type formulae are modified by vacuum polarization,
\begin{equation}
  \delta_{h, n} \rightarrow \delta_{h, n}+ \delta_{h, n}^{\rm vac}, \quad \delta_{h, n}' \rightarrow \delta_{h, n}'+ \delta_{h, n}^{\rm vac},
\end{equation}
where
\begin{equation}
  \delta^{\rm vac}_{h, n}=\frac{2\delta \psi_{h, n}(0)}{\psi_{h, n}(0)}.
\end{equation}
Formally, the contribution $\delta^{\rm vac}_{h, n}$ is of order $\alpha^2$, but
  enhanced because of the large coefficient containing $\mu_+/m_e$. For the
  ground state, the corrections \cite{Eiras:2000rh} yield $\delta^{\rm vac}_{{\sss K}, 1}=0.45\cdot 10^{-2}$ and $\delta^{\rm
  vac}_{\pi, 1}=0.31\cdot 10^{-2}$. The changes in $\delta_{\pi, 1}$ and
  $\delta_{\pi, 1}'$ due to $\delta^{\rm vac}_{\pi, 1}$ are about $5\%$. For
  $\delta_{{\sss K}, 1}'$ however, the correction amounts to $27\%$. Here, we omit
  the contributions from $\delta_{h, n}^{\rm vac}$, because the uncertainties
  in $\delta_{h, n}$ and $\delta_{h, n}'$ are much larger than $\delta^{\rm vac}_{h, n}$.
\begin{table}[t]
\begin{center}
\begin{tabular}{|l|r|r|r|r|r|}\hline
$\pi^\pm K^\mp$ atom&$\Delta E^{\rm em}_{nl}$[eV]&$\Delta E^{\rm vac}_{nl}$[eV]&
$\Delta E^{\rm h}_{n}$[eV]&
$\tau_{n}$[s]\\\hline
$n$=1, $l$=0&$-0.095$&$-2.56$&$-9.0\pm1.1$&$(3.7\pm0.4)\cdot10^{-15}$\\
$n$=2, $l$=0&$-0.019$&$-0.29$&$-1.1\pm0.1$&\\
$n$=2, $l$=1&$-0.006$&$-0.02$&&\\\hline
\end{tabular}
\end{center}
 \medskip
 \caption{Numerical values for the energy shift and the lifetime of the $\pi^\pm K^\mp$ atom.\label{table: numericsK}}
\end{table}
\begin{table}[t]
\begin{center}
\begin{tabular}{|l|r|r|r|r|r}\hline
$\pi^+\pi^-$ atom & $\Delta E^{\rm em}_{\pi, nl}$[eV]&$\Delta E^{\rm vac}_{\pi,
  nl}$[eV]&$\Delta E^{\rm h}_{\pi, n}$[eV]&$\tau_{\pi, n}$[s]\\\hline
$n$=1, $l$=0&$-0.065$&$-0.942$&$-3.8\pm 0.1$ &$(2.9\pm0.1)\cdot10^{-15}$\\
$n$=2, $l$=0&$-0.012$&$-0.111$&$-0.47\pm 0.01$&\\
$n$=2, $l$=1&$-0.004$&$-0.004$&&\\\hline
\end{tabular}
\end{center}
 \medskip
 \caption{Numerical values for the energy shift and the lifetime of the
   $\pi^+\pi^-$ atom.\label{table: numericspi}}
\end{table}

The numerical values for the lifetime $\tau_1 \doteq \Gamma^{-1}_1$,
($\tau_{\pi, 1} \doteq \Gamma^{-1}_{\pi, 1}$) and the energy shifts at
next-to-leading order in isospin symmetry breaking are given in
table \ref{table: numericsK} and \ref{table: numericspi}. The energy shifts due to vacuum polarization
$\Delta E^{\rm vac}_{nl}$ are taken from Ref. \cite{Eiras:2000rh,vacpol}. In the
evaluation of the uncertainties, the correlations between the S-wave scattering lengths have been
taken into account. For the decay
width and the strong energy shift of the $\pi^\pm K^\mp$ atom, the dominant
source of uncertainty is due to the uncertainties in the scattering lengths $a_0^+$ and $a_0^-$. We do not display the error bars for the electromagnetic
energy shifts, which stem at order $\alpha^4$ from the uncertainties in
$\langle r^2_{\pi^+}\rangle$ and $\langle r^2_{\sss
  K^+}\rangle$ only. For pionium, the uncertainties of $\Delta E_{\pi, 10}^{\rm em}$
at order $\alpha^4$ amount to about $0.7\%$, while for the $\pi^\pm K^\mp$
atom  $\Delta E_{10}^{\rm em}$ is known at the $5\%$ level. To estimate the order of magnitude of the electromagnetic corrections at higher
order, we may compare with positronium. Here, the $\alpha^5$ and $\alpha^5
\,{\rm ln}\alpha$ corrections \cite{Karplus:1952wp} amount to about $2\%$ with
respect to the $\alpha^4$ contributions.
\section{Summary and Conclusions}
We provided the formulae for the energy shifts and decay widths of the
$\pi^+\pi^-$ and $\pi^\pm K^\mp$ atoms at next-to-leading order in isospin
symmetry breaking. To confront these predictions with data presents a
challenge for future hadronic atom experiments. Should it turn out that these
predictions are in conflict with experiment, one would have to revise our
present understanding of the low-energy structure of QCD. 
 \section*{Acknowledgments}
  It is a great pleasure to thank J. Gasser for many interesting discussions and
  suggestions. Further, I thank R. Kaiser, A. Rusetsky, H. Sazdjian and
  J. Schacher for useful discussions as well as for helpful comments on the manuscript. This work was supported in part by the Swiss National Science Foundation and by
 RTN, BBW-Contract N0.~01.0357 and EC-Contract HPRN--CT2002--00311 (EURIDICE).

\end{document}